\documentclass{webofc}
\usepackage[varg]{txfonts}   
\usepackage{gensymb}

\usepackage{natbib,hyperref}
\usepackage{siunitx}
\usepackage{indentfirst}
\usepackage{makecell}

\begin{document}
\title{Probing the subsurface of the two faces of Iapetus}
%
%

\author{\firstname{L\'ea E.} \lastname{Bonnefoy}\inst{1,2}\fnsep\thanks{\email{lea.bonnefoy@latmos.ipsl.fr}} \and 
	\firstname{Jean-Fran\c cois} \lastname{Lestrade}\inst{3} \and
	\firstname{Emmanuel} \lastname{Lellouch}\inst{2} \and
	\firstname{Alice} \lastname{Le Gall}\inst{1}  \and
	\firstname{C\'edric} \lastname{Leyrat}\inst{2}  \and
	\firstname{Nicolas} \lastname{Ponthieu}\inst{4}  \and
	\firstname{Bilal} \lastname{Ladjelate}\inst{5} 
}

\institute{Laboratoire Atmosphères, Milieux, Observations Spatiales (LATMOS), UVSQ /CNRS/Paris VI, Guyancourt, France  
\and
           Laboratoire d'Etudes Spatiales et d'Instrumentation en Astrophysique (LESIA), Observatoire de Paris-Meudon, Meudon, France 
\and
           LERMA, Observatoire de Paris, PSL Research University,CNRS, Sorbonne Universit\'es, UPMC Univ. Paris 06, 75014 Paris, France  
\and
		Univ. Grenoble Alpes, CNRS, IPAG, Grenoble, France 
\and
		Institut de Radioastronomie Millimétrique, Granada, España
          }

\abstract{%
  Saturn's moon Iapetus, which is in synchronous rotation, is covered by an optically dark material mainly on its leading side, while its trailing side is significantly brighter. Because longer wavelengths probe deeper into the subsurface, observing both sides at a variety of wavelengths brings to light possible changes in thermal, compositional, and physical properties with depth. We have observed Iapetus's leading and trailing hemispheres at 1.2 and 2.0 mm, using the NIKA2 camera mounted on the IRAM 30-m telescope, and compared our observations to others performed at mm to cm wavelengths. We calibrate our observations on Titan, which is simultaneously observed within the field of view. Due to the proximity of Saturn, it is sometimes difficult to separate Iapetus's and Titan's flux from that of Saturn, detected in the telescope's side lobes. Preliminary results show that the trailing hemisphere brightness temperatures at the two wavelengths are equal within error bars, unlike the prediction made by Ries (2012)\cite{ries2012}. On the leading side, we report a steep spectral slope of increasing brightness temperature (by 10 K) from 1.2 to 2.0 mm, which may indicate rapidly varying emissivities within the top few centimeters of the surface. Comparison to a diffuse scattering model and a thermal model will be necessary to further constrain the thermophysical properties of the subsurface of Iapetus's two faces.  
}
\maketitle
\section{Introduction}
\label{intro}

Saturn’s icy satellites, which are in synchronous rotation around Saturn, often display large differences between their leading and trailing sides, which interact differently with their orbital environment and, in particular, with Saturn's dust rings. This is especially true for Iapetus, which presents the largest albedo dichotomy in the Solar System. Indeed, a dark material covers the leading side as it travels through the diffuse Phoebe ring \cite{verbiscer2009}, and thermal segregation further enhances the resulting albedo contrast \cite{spencer2010}. This dichotomy, obvious in the visible, has also been observed by the Cassini spacecraft in the far-infrared with the Composite Infrared Spectrometer (CIRS, \SI{7}{\um}–\SI{1}{\mm}) \cite{howett2010} and at 2.2 cm with the Cassini Radar/Radiometer \cite{ostro2006,ostro2010,legall2014}. The Arecibo 12.6-cm Radar, however, showed no asymmetry between leading and trailing, indicating that the dark layer covering the leading side is most likely less than a meter thick \cite{black2004}.

Ries (2012) partially bridged the gap between CIRS and Cassini radiometry by observing Iapetus’s two faces at wavelengths varying from 3 to 10.8 mm using the Green Bank Telescope (GBT) \cite{ries2012}. He observed that, while the trailing side is less emissive than the leading side, it also shows a large absorption feature likely centered near 3 mm. He attributed this feature to diffuse scattering by 1–2-mm ice particles, by comparison with the semi-empirical Microwave Emission Model for Layered Snowpacks (MEMLS) developed for and tested on snow on Earth \cite{wiesmann1998}. He predicted that observations below 3 mm should show a progressive drop of the emissivity with increasing wavelength. 

To complete a missing part of Iapetus’s microwave spectrum, we observed the two faces of Iapetus at 1.2 and 2.0 mm using the NIKA2 \cite{catalano2014,adam2018,calvo2016} at the IRAM-30 m telescope at Pico Veleta in Spain. 



\section{Observations}
\label{sec-2} 

\begin{figure}[h]
	\begin{center}
		\includegraphics[width=10cm]{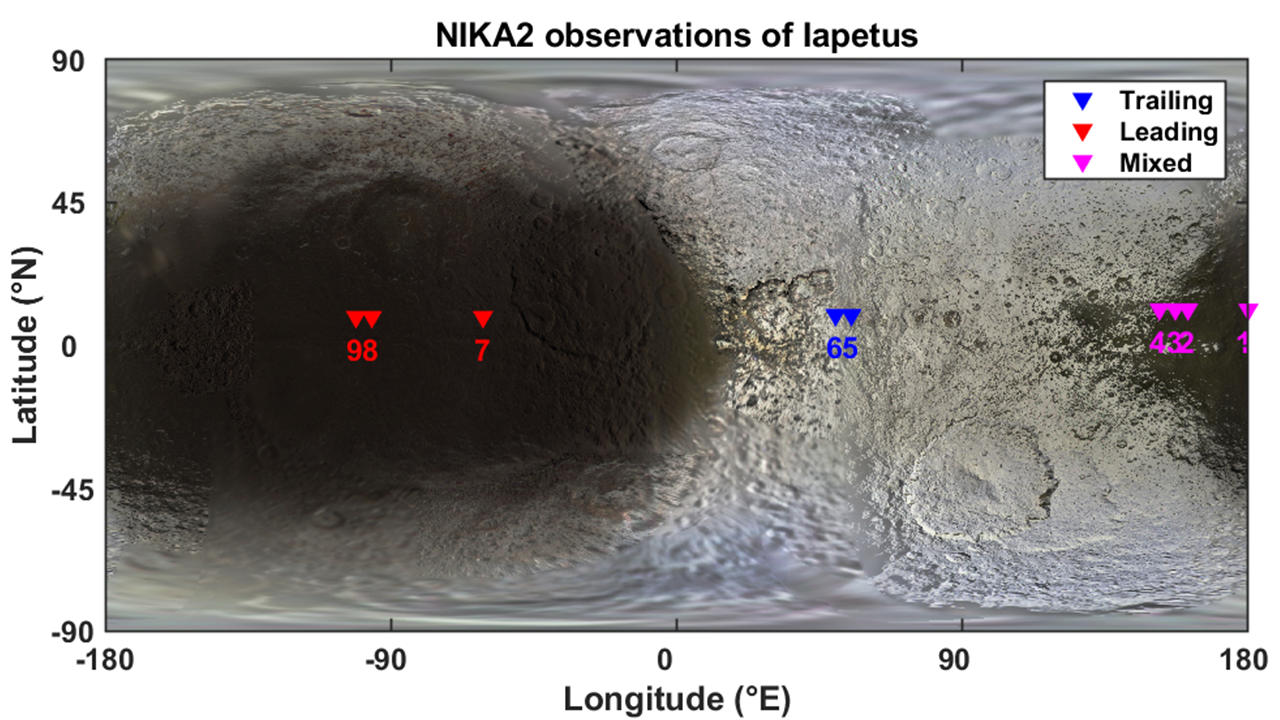}
		\caption{Map of Iapetus with the sub-Earth point indicated for each date of observation. The observations are numbered for reference in Table.~\ref{fig-1}. The background image is a mosaic of Cassini UV and optical data (credit: NASA/JPL-Caltech/SSI/Lunar and Planetary Institute)}
		\label{fig-1}       
	\end{center}
\end{figure}

\begin{table}
	\centering
	\caption{NIKA2 observations of Iapetus. The latitude and longitude given are for the sub-Earth point on Iapetus. The sky opacity $\tau$ at zenith is measured with a 225 GHz (1.3 mm) taumeter. The observations are numbered for reference in Fig.~\ref{fig-1}. }
	\label{tab-1}       
	\begin{tabular}{llllllllll}
		\hline
		\thead{\#\\~} & \thead{Date\\~} & \thead{Elevation\\~} & \thead{$\tau$\\~} & \thead{Lat.\\(\degree N)} & \thead{Long.\\(\degree E)} & \thead{Iapetus\\elong.} & \thead{Titan\\elong.} & \thead{Int.\\time} & \thead{Observed\\side} \\ 
		1 & 23 May 2018 & 27.6-30.7	& 0.28 & 11.2 & 179 & 95" & 186" & 1.5h & Mixed \\
		2 & 27 May 2018 & 28.7-30.4 & 0.56 & 11.3 & 161 & 212" & 106" & 0.7h & Mixed \\
		3 & 28 May 2018 & 22.2-30.7 & 0.21 & 11.3 & 157 & 250" & 153" & 3.9h & Mixed \\
		4 & 29 May 2018 & 21.5-30.7 & 0.20 & 11.3 & 152 & 291" & 185" & 1.5h & Mixed \\
		5 & 14 Feb 2019 & 27.4-31.0 & 0.13 & 9.5 & 55 & 432" & 92" & 2.4h &  Trailing\\
		6 & 15 Feb 2019 & 28.1-30.9 & 0.21 & 9.5 & 50 & 412" & 132" & 2.4h & Trailing \\
		7 & 12 Mar 2019 & 29.7-31.3 & 0.10 & 8.9 & -61 & 412" & 156" & 1.8h & Leading \\
		8 & 20 Mar 2019 & 20.2-31.3 & 0.13 & 8.8 & -96 & 495" & 163" & 2.4h & Leading \\
		9 & 21 Mar 2019 & 28.5-31.3 & 0.17 & 8.8 & -101 & 491" & 179" & 2.4h & Leading \\
	\end{tabular}
\end{table}

In order to measure the flux densities of the trailing and leading hemispheres of Iapetus independently, we had to observe the satellite near maximum elongation. Given that Iapetus orbits around Saturn in 79 days, the most appropriate dates to observe during the winter 2019 pools were January 31 to February 15 for the trailing side, and March 12 to March 26 for the leading side. We also observed a mix of the leading and trailing sides in May 2018; the observations of May 28 and 29, 2018 will be integrated in our final results once they are calibrated. While calibrators such as Uranus were observed before or after each Iapetus observation, it is preferable to calibrate the data on Titan. Indeed, Titan's brightness temperature spectrum is accurately known (<5\% uncertainty \cite{lellouch2019}), and more importantly it was observed at the same time and under the same atmospheric conditions as Iapetus, owing to the large field of view of NIKA2 and the map sizes. This imposes further timing constraints for the observations, as it is only possible to accurately measure the flux from Titan near its maximum elongation, when it is more easily separable from Saturn.

Using the NIKA2 millimeter camera, we imaged a field of view centered on Saturn and extending slightly beyond the position of Iapetus, 200” to 500” away from Saturn depending on the date. The characteristics of the Iapetus observations are summarized in Table~\ref{tab-1}, and their sub-Earth longitudes are shown on a map of Iapetus in Fig.~\ref{fig-1}. Note that on May 23, 2018 Iapetus was too close to Saturn for an accurate determination of its flux; on May 27 the opacity was too high to detect Iapetus; and on February 14, 2019 Titan was too close to Saturn for calibration.

\setlength{\parskip}{0em}
\section{Methods}
\label{sec-3}

\begin{figure}[h]
	\begin{center}
		\includegraphics[width=12cm]{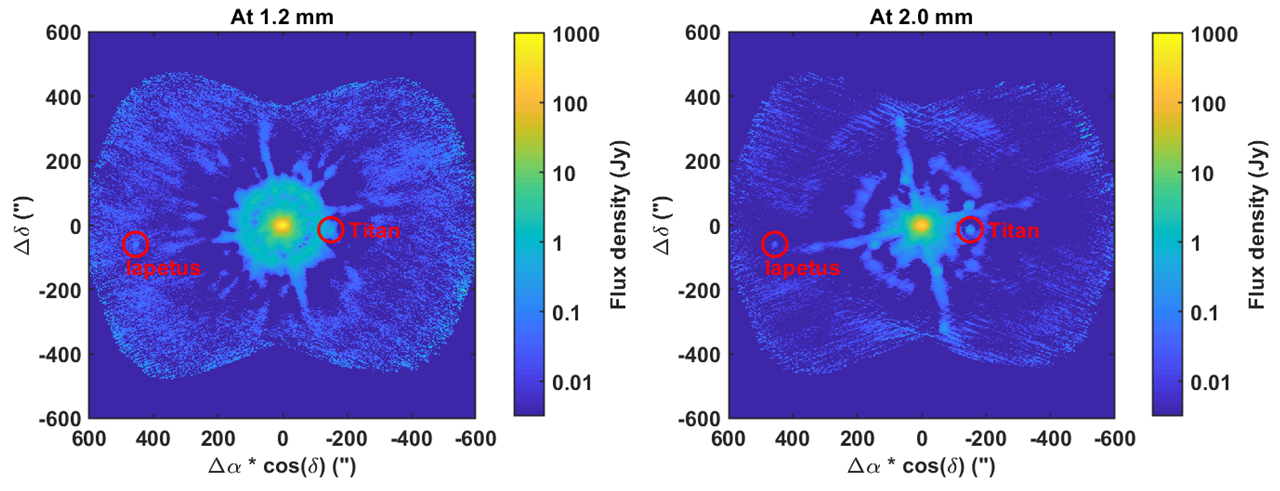}
		\caption{Observation of Saturn and its satellites on 20 March, 2019. Data is integrated over 30 minutes, from 4:58 to 5:40 UTC. The positions of Titan and Iapetus are indicated; Saturn is in the center. Other rings and spots up to $\sim$400"  from the center reflect the extended beam pattern of the IRAM 30 m telescope and NIKA2, visible because Saturn is a very bright source (cf beam pattern shown in Adam et al., 2018 \cite{adam2018}). The other mid-sized satellites of Saturn are too close to Saturn to separate. }
		\label{fig-2}       
	\end{center}
\end{figure}

\begin{figure}[h] 
	\begin{center}
		\includegraphics[width=12.5cm]{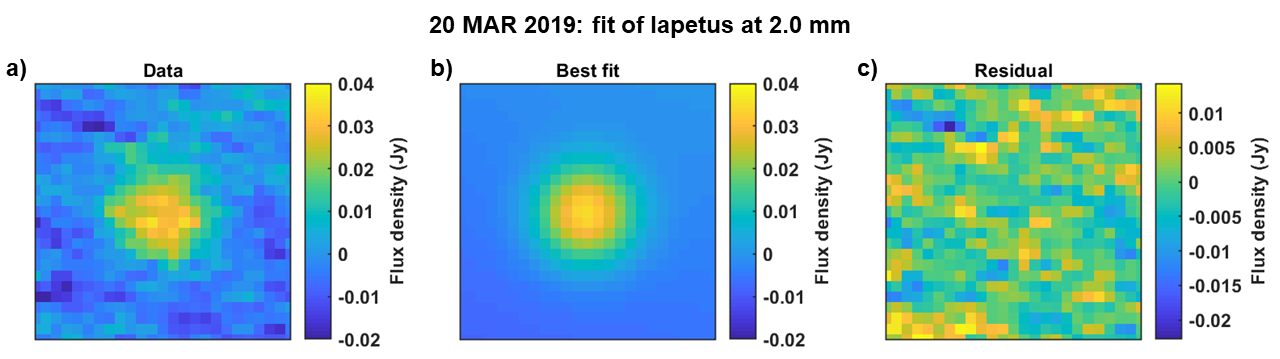}
		\caption{a) View of Iapetus on 20 March, 2019. Data is integrated over 30 minutes, from 4:58 to 5:40 UTC. Image is shown for the 25x25" window where the fit is applied at 2.0 mm. Note this is the same data shown in Fig.~\ref{fig-2}b. b) Best fit to the data. The model fitted is a Gaussian over a tilted plane; in this case, the Iapetus flux found is $0.039\pm 0.002$ Jy. c) Residual map, with a mean of 0 and a standard deviation of 0.005 Jy. }
		\label{fig-4}       
	\end{center}
\end{figure}

The measurement of both Titan’s and Iapetus’s flux density is complicated by the proximity of Saturn which, at $\sim$1200 Jy at 1.2 mm and $\sim$500 Jy at 2.0 mm, is over 10 000 times brighter than Iapetus ($\sim$0.1 Jy at 1.2 mm and $\sim$0.03 Jy at 2.0 mm) and 1000 times brighter than Titan ($\sim$1.2 Jy at 1.2 mm and $\sim$0.4 Jy at 2.0 mm) at mm wavelengths. The large-scale structures, such as the error beam \cite{kramer2013}, are filtered out along with the atmospheric fluctuations in the data reduction pipeline \cite{ruppin2018}. However, residual structures remain: these are visible in Fig.~\ref{fig-2}, and affect the data out to at least 350" from Saturn. We account for the extended beam pattern using the following method.

As the telescope follows Saturn in the sky over the 1 to 4 hours of observation, the parallactic angle changes. In practice, this means that the orientation of the beam pattern changes relative to the position of the satellites. Thus over the hours of observation, both Titan and Iapetus move in and out of the peaks within the sidelobes. Our method consists in dividing our observations into 30-minute segments, which is a time span short enough that the position of the satellites within the beam varies little, and long enough that we have a clear detection. The positions of Saturn, Titan, and Iapetus are known precisely at each time using the SPICE toolkit developed by tha Navigation and Ancillary Information Facility (NAIF) \cite{acton2018}. For each 30-minute segment, we select the data over a small region (19x19” at 1mm; 25x25” at 2mm) centered on the satellite; these sizes have been optimised to properly separate the Gaussian parameters and the background. Within this small region, we fit a 2-dimensional Gaussian with a tilted plane in the background; the amplitude of the Gaussian is the satellite's flux. The tilted plane accounts for variations in the background flux caused by Saturn flux in the extended beam pattern and by the negative rebound introduced by the atmosphere subtraction. An example of this method is shown in Fig.~\ref{fig-4}. 

\par Once the Titan and Iapetus fluxes are found, we perform an absolute calibration on Titan, using the spectrum from Lellouch et al. (2019) \cite{lellouch2019}, then convert the Iapetus flux to brightness temperature (the angular size of Iapetus is around 0.2" and varies with the Earth-Saturn distance). Individual values for each segment are shown in Fig.~\ref{fig-3}a. Error bars are derived from the 95\% confidence interval of the 2-D Gaussian fits to the fluxes of Iapetus and Titan; residual errors due to the extended beam are most likely the cause of the observed scatter within each day (see Fig.~\ref{fig-3}a). Estimating the uncertainty on each fit with a Monte Carlo method yields very similar error bars, except when the residual noise is not Gaussian, in which case a Monte Carlo method underestimates the uncertainty. For the leading side, all 1.2-mm and 2.0-mm are consistent from day to day, so we average the total of 11 segments from March 12, 20, and 21, 2019. Over the hours and days of the March observations, the position of Titan and Iapetus relative to the extended beam pattern varies enough that we can safely assume that the influence of the sidelobes averages out. For the trailing side, Titan is too close to Saturn on February 14, causing the large scatter in the calibrated data; we therefore only average the 4 segments of February 15, 2019. Because there are only four averaged points and they are all from the same day when Titan was still relatively close to Saturn, there may be an offset due to the calibration on Titan. The leading and trailing values are shown in Fig.~\ref{fig-3}b; error bars are propagated from the uncertainty on individual measurements shown in Fig.~\ref{fig-3}a while also taking into account the 5\% uncertainty on the Titan model. Using instead the variance of the data yields similar but slightly smaller uncertainties. 


 \section{Results and interpretations}
\label{sec-4}

\begin{figure}[h]
	\begin{center}
		\includegraphics[width=12.5cm]{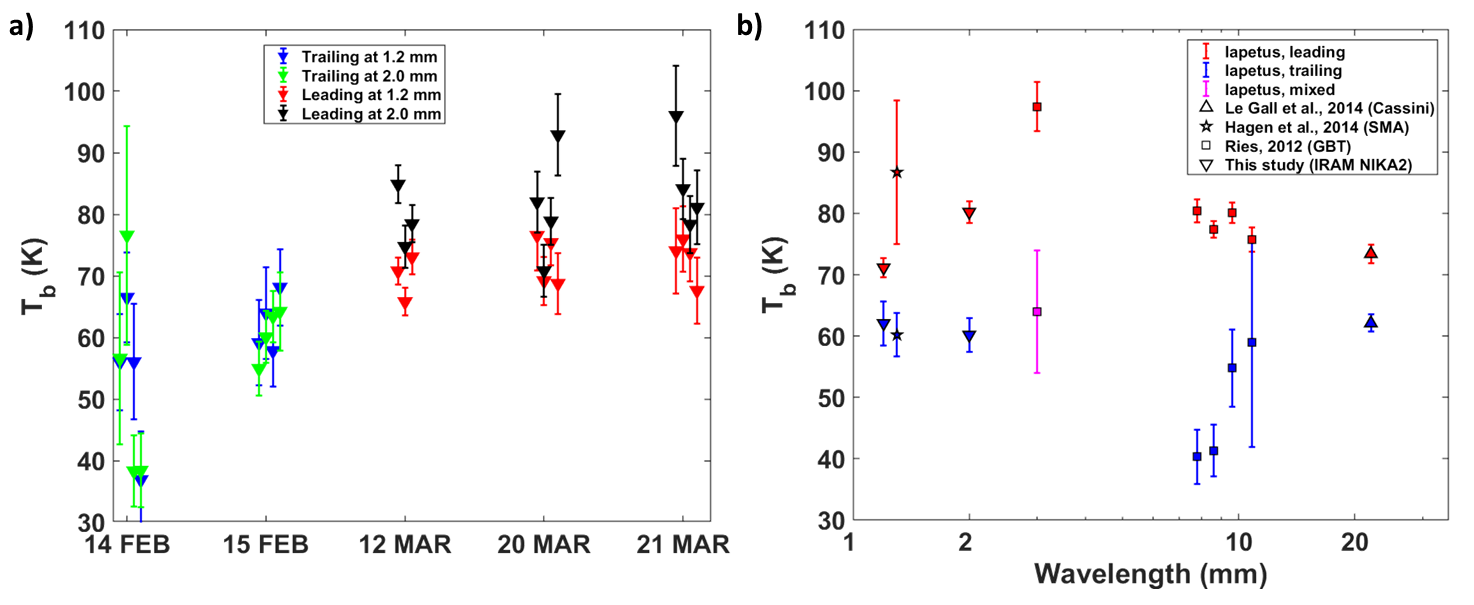}
		\caption{a) Measured brightness temperatures for every 30-minute segment of every day in 2019, at 1.2 and 2.0 mm. Error bars are derived from the 95\% confidence interval of the 2-D Gaussian fits to the fluxes of Iapetus and Titan; errors due to the extended beam are most likely the cause of the observed scatter. The 5\% uncertainty on the Titan flux is not included in the error bars as it would shift all data in the same direction, without changing relative values. b) Iapetus microwave spectrum, including the 1.2- and 2.0-mm values found with the NIKA2 camera as well as brightness temperatures found in previous studies \cite{ries2012,hagen2014,legall2014}. Our error bars are derived from the 5\% Titan flux uncertainty and the 95\% confidence interval of the fits, as detailed in Section \ref{sec-3}. }
		\label{fig-3}       
	\end{center}
\end{figure}

The preliminary 1.2- and 2.0-mm brightness temperatures are shown in Fig.~\ref{fig-3}b alongside those found in previous studies \cite{ries2012,hagen2014,legall2014}. As expected, the leading side, which is covered by the optically dark material, has significantly higher brightness temperatures than the trailing side. 

On the trailing side, the 1.2-mm, 1.3-mm (from the SubMillimeter Array (SMA),\cite{hagen2014}), and 2.0-mm brightness temperatures are all similar within error bars. While they do not follow the decreasing trend predicted by Ries (2012), they are not incompatible with his interpretations. Indeed, it remains likely that the absorbing feature visible in the GBT data is due to diffuse scattering by ice particles on this hemisphere which has a composition dominated by water ice. We also highlight that the calibration of the data collected on the trailing side on February 14 and 15, 2019 can be improved. At those dates and especially on the 14th, Titan's elongation was relatively small and the calibration is more strongly affected by Saturn flux leaking into the sidelobes. Improved data analysis methods such as subtracting a beam pattern or calibrating on Saturn will help constrain the trailing brightness temperature. 

On the leading side, we observe a very steep spectral slope from \SI{1} to \SI{3}{\mm}. Either this steep spectral slope is a property intrinsic to the optically dark material covering the leading side, or it indicates that the subsurface properties change very quickly with depth in the top few cm of the subsurface. 

\section{Conclusion}
\label{sec-5}

The brightness difference between the leading and trailing sides of Iapetus is visible at all millimeter wavelengths. Preliminary IRAM NIKA2 1.2- and 2.0-mm results show a steep slope on the leading, but remain equal within error bars on the trailing, unlike what was predicted by Ries (2012) \cite{ries2012}. Further analysis will use the new version of the NIKA2 data reduction pipeline, and will include subtraction of the extended beam pattern, especially to improve the quality of the May 2018 and February 2019 data. Using a model of Saturn and its rings, we are also planning to compare calibration on Titan and on Saturn. 



%
%
%
\bibliography{thebib}
%
%

\end{document}